\documentclass[journal,12pt,draftclsnofoot,onecolumn]{IEEEtran}
\usepackage{amsmath,amssymb,amsfonts,mathrsfs}
\usepackage{algorithmic}
\usepackage{graphicx}
\usepackage{textcomp}
\usepackage{braket}
\usepackage{soul}
\usepackage{xcolor}
\usepackage{subfig}

\ifCLASSINFOpdf
\else
\fi
\begin{document}
%
\title{Energy Optimization \\ for Time-of-Arrival Based Tracking}
%
%
%

\author{Luca~Reggiani, Arnaldo~Spalvieri
\thanks{L. Reggiani and A. Spalvieri are with Dipartimento di Elettronica, Informazione e Bioingegneria, Politecnico di Milano, Italy (email: luca.reggiani@polimi.it, arnaldo.spalvieri@polimi.it).}
}

\maketitle

\begin{abstract}
The paper analyzes energy allocation in a scenario where the position of a moving target is tracked by exploiting the Time-of-Arrivals of bandwidth-constrained signals received by or transmitted from a fixed number of anchors located at known positions. 
The signal of each anchor is generated by transmitting a sequence of known symbols, allowing for amplitude and duration (number of symbols) to be different from anchor to anchor. The problem is the minimization of the sum of the energies of the transmitted signals imposing a constraint on the performance of the tracking procedure. Specifically, the constraint is the Posterior Cramer-Rao Bound, below the mean square error achieved by any unbiased estimator.  
The main improvement over the previous literature is the derivation of a formula that, at each step of the tracking, allows to calculate in closed form the first-order variation of the Posterior Cramer-Rao Bound as a function of the variation of the total energy. To concretely show the application of our approach, we present also two numerical algorithms that implement the constrained optimization in the case of signals of fixed amplitude and variable duration transmitted from the anchors in a time division multiplexing scheme. 
\end{abstract}


%
\IEEEpeerreviewmaketitle

\section{Introduction}

\IEEEPARstart{L}{ocalization} is a fundamental task in several indoor and outdoor systems, that can become crucial in certain applications as the industrial and the vehicular ones. The latest 5-th generation of mobile communication standards and technologies poses demanding performance targets, expressed in terms of increased precision and availability, reduced energy, latency and robustness, thanks also to higher and wider frequency bands, dense cells deployment and renewed waveform design.
%
%
The technical solutions proposed for 5G New Radio (NR) positioning are covering different approaches and measurement principles, from Time of Arrival (ToA) to Angle of Arrival AoA or Angle of Departure (AoD) \cite{10127621, 9522056}.
The 3GPP 5G radio-technology provides enhanced location capabilities starting from Release 16, focusing on advanced capabilities like signal design, antenna configuration or network coordination, and the implementation of highly precise positioning functions has reached further maturity in Release 17, frozen in 2022 \cite{3GPPtr21917}. 

Due to the strict energy requirements, the issue of achieving accuracy in the estimate of target position at low energy cost arises in virtually all the applications involving Wireless Sensor Networks (WSN) and IoT devices. We observe that, when the energy necessary for achieving the desired accuracy is reached by transmitting sequences of a variable number of impulses each of which has the same energy, the energy becomes proportional to the latency, hence energy minimization becomes equivalent to latency minimization. The issue of latency minimization typically arises in those applications that need to track target position along time, including certain industrial applications where the control of manufactures, robots and other machines needs both high accuracy, velocity, and fast control loops. Similar requirements hold also in the case of automated vehicles, including the aerial ones. The importance of the trade off between latency and positioning accuracy is confirmed by the recent evolutions in the standardization processes, see for instance  \cite{3GPPtr22804}, where a study on communication for automation in virtual domain has identified a set of performance targets for industrial automation using 5G. In certain cases, the latency allowable over the network may constrain the transmission time to be as low as $500$ $\mu s$ \cite{brown2018ultra}.
Relevant examples of performance requirements can be found also in \cite{3GPPtr38857}, where it is reported a wide study on Release 17 of 3GPP 5G \cite{3GPPtr21917} target positioning requirements for commercial use cases and IIoT (Industrial Internet of Things) use cases. Accuracy is investigated in its components, vertical and horizontal, latency is defined for the uplink and/or downlink, and it is also introduced the definition of the integrity of the positioning estimate in order to define and measure its reliability. The target position requirements for the most demanding applications are in the decimetre range ($< 0.2$ m) for horizontal accuracy for $90\%$ of devices, less than $1$ m for vertical accuracy and $90\%$ of devices, the physical layer latency less than $10$ ms and the availability of reliable measures from $95\%$ to $99.9\%$ and more.

These considerations motivate the study of the trade-off between the energy spent at the physical layer for tracking and the accuracy of the estimated position that the tracking procedure can achieve. To do this, we exploit the direct relation between the quality of the ranging measure, the Signal-to-Noise Ratio (SNR), and the energy spent for the transmission of the ranging signals. 
We mainly focus on the downlink, where the anchors transmit the ranging signals, e.g. in a time division multiplexing scheme, and the moving target device receives and analyzes them. In this scenario, the target device is connected by a data channel to the anchors and, through this channel, after the analysis communicates to the anchors how to adjust the transmission parameters, even if the energy spent for this channel is not considered in this paper. Specifically, the adjustment is the result of a constrained optimization, where the energy spent by the anchors for transmission at the next step of tracking is minimized under a constraint on the Posterior Cramer-Rao Bound (PCRB) on the achievable accuracy of the current estimate of target position. By reciprocity,  when the target transmits different signals to the different anchors, the analysis of the uplink is the same as that of the downlink, but, at the system level, here the signals detected by the anchors must be made available to a centralized data processing unit that analyzes them. When, instead, the target transmits only one signal that is received from all the anchors, the analysis is based on a straightforward modification of the analysis made for the downlink.

The paper is organized as follows. After a review of the literature that is more strictly related to our work, in Sect. \ref{Sec.:Related_works} we outline our main new contributions. Sect. \ref{Sec.:System Model} presents the system and signal models considered for the analysis. Sect. \ref{Sec:Recursive PCRB} revises the (recursive) PCRB below the achievable Mean-Square Error (MSE) between the actual target position and the estimated one along the tracking process. Our main contributions are presented in Sect. \ref{Sec.:Problem Formulation}, where, after the formulation of the problem, in Subsection \ref{subsec:First_order_variation_PCRB} we give closed-form formulas for the variation of the positioning error as a function of the variation in the transmitted energy. Based on these formulas, in Subsections \ref{subsec:delta_algorithm} and \ref{subsec:sliding_space}, we propose two concrete tracking algorithms, showing in Subsection \ref{subsec:sliding_space_benchmark} how they apply to latency optimization. Numerical results are reported and commented on in Sect. \ref{sec:numerical_results}, while the conclusions are drawn in Sect. \ref{sec:conclusions}.

\section{Related works and novelties of this paper}
\label{Sec.:Related_works}

There is a vast literature, especially for wireless sensor networks, that investigates the trade-off between energy efficiency and accuracy of tracking procedures. This trade-off, which plays a crucial role in several applications, is a hot topic that is attracting increasing interest today, as demonstrated by the large number of recent papers on the subject; for example, see the very recent \cite{10620268}. 
In \cite{6196144}, a review on energy efficiency in collaborative target tracking applications has been presented for the relevant case of WSNs: the numerous aspects related to achieving high energy efficiency in tracking applications are divided into sensing-related and communication-related ones. In the former category, information-driven techniques exploit the data content to optimize future readings, while data filtering techniques generate information from noisy readings. In the latter category, node selection, sleep scheduling, and dynamic
clustering are considered as the main techniques to optimize energy consumption during tracking / localization processes. Data filtering techniques is the subcategory that is closer to the contribution of this paper and, in particular, the contributions that investigate modifications and integrations of the Extended Kalman Filter (EKF), with the aim of increasing the energy efficiency of the collaborative tracking. In this context, this work aims to provide additional insight into the relation between a variation in energy allocation and the theoretical impact on the PCRB, which can also be translated into the implementation of the EKF in a tracking process.


In \cite{PANG2021932}, in addition to the optimization of the sensor management scheme (i.e., how to select nodes in the WSN), which is common in many papers in the energy efficiency area, the time interval between two observations and the selection of useful observations are optimized within a distributed Kalman filter. 
More recently, in \cite{Hchen_10049506,Li_9881913,Huang_10824973,9119156}, the optimization of resource allocation has been faced in the context of a Multiple Input Multiple Output radar for tracking multiple targets. Specifically, in \cite{9119156} the PCRB is used for formulating the optimization problem: this is analogous to the approach presented here, and the optimization method is obtained by combining convex relaxation and the cyclical minimizer for the specific application with a single, multi-antenna anchor.



Other papers are related to this work, even if they focus on the latency instead of the energy trade-off; however, the two approaches can be strictly related in some cases of interest (Sect. \ref{subsec:alg_latency_power}). In \cite{7346955} a latency constrained optimization algorithm is investigated in the context of a received signal strength based algorithm for estimating a target location through the visible light channel; the crucial contribution of this work has been the formalization of a trade-off between accuracy and latency, although the formalization applies only within the specific boundaries of the considered scenario. 
In \cite{Haojun_2019}, it is analyzed the impact of the duration of the observation on the positioning accuracy for Bluetooth low energy based fingerprint positioning: the Cramér-Rao Bound (CRB) of localization using the received signal strength is elaborated for revealing the relation between delay and accuracy and providing a reference for an optimal selection of the window size in the positioning process.
Finally, in \cite{6781631} positioning accuracy in UWB networks is investigated in terms of trade-off with the Medium Access Control (MAC) delay, showing that the full cooperative approach aiming at the accuracy maximization usually brings to significant delays. Therefore a correct trade-off and appropriate strategies are necessary in practical applications.
%

%
Compared with the cited literature, the main novel contributions of this paper are the following:
\begin{enumerate}
	\item The main contribution of the paper is a closed-form formula that expresses the first-order variation of the PCRB of target position w.r.t. the variation of the transmitted energy at the anchors. This formula dramatically reduces the complexity of the constrained optimization problem, which would elsewhere require importing the entire function that represents the PCRB in the Karush-Kuhn-Tucker (KKT) conditions. To this regard, we observe that also in \cite{6807568} the KKT conditions are imposed, but the constrained optimization problem analyzed in \cite{6807568} is the less demanding one of position estimation, not tracking. The approach that we adopt to derive this formula is more general
 than the approach of \cite{7346955}. As such, the proposed approach can be applied to different optimization algorithms and to various transmission parameters, including latency.
	\item Based on this theoretical analysis, formulation of the optimization problem for minimizing the energy or the latency (this extends the approach of \cite{9119156}, which is limited to the optimization of the amplitude of the signal), given a target PCRB for downlink or uplink scenarios. 
	\item Implementation and simulation of two algorithms for the constrained minimization. The first algorithm is a \textit{constrained gradient descent method} that calculates the gradient from the formula that relates energy variation and accuracy. The second algorithm is more computationally demanding, as it performs an exhaustive search 
    in the neighborhood of the current estimate of the position in place of computing the gradient. The two algorithms are compared to a brute-force approach not exploiting the PCRB variation analysis, in order to validate it and provide a measure of the potential energy savings.
\end{enumerate}

\begin{figure}[!tb]
	\centering
	\includegraphics[scale=0.5]{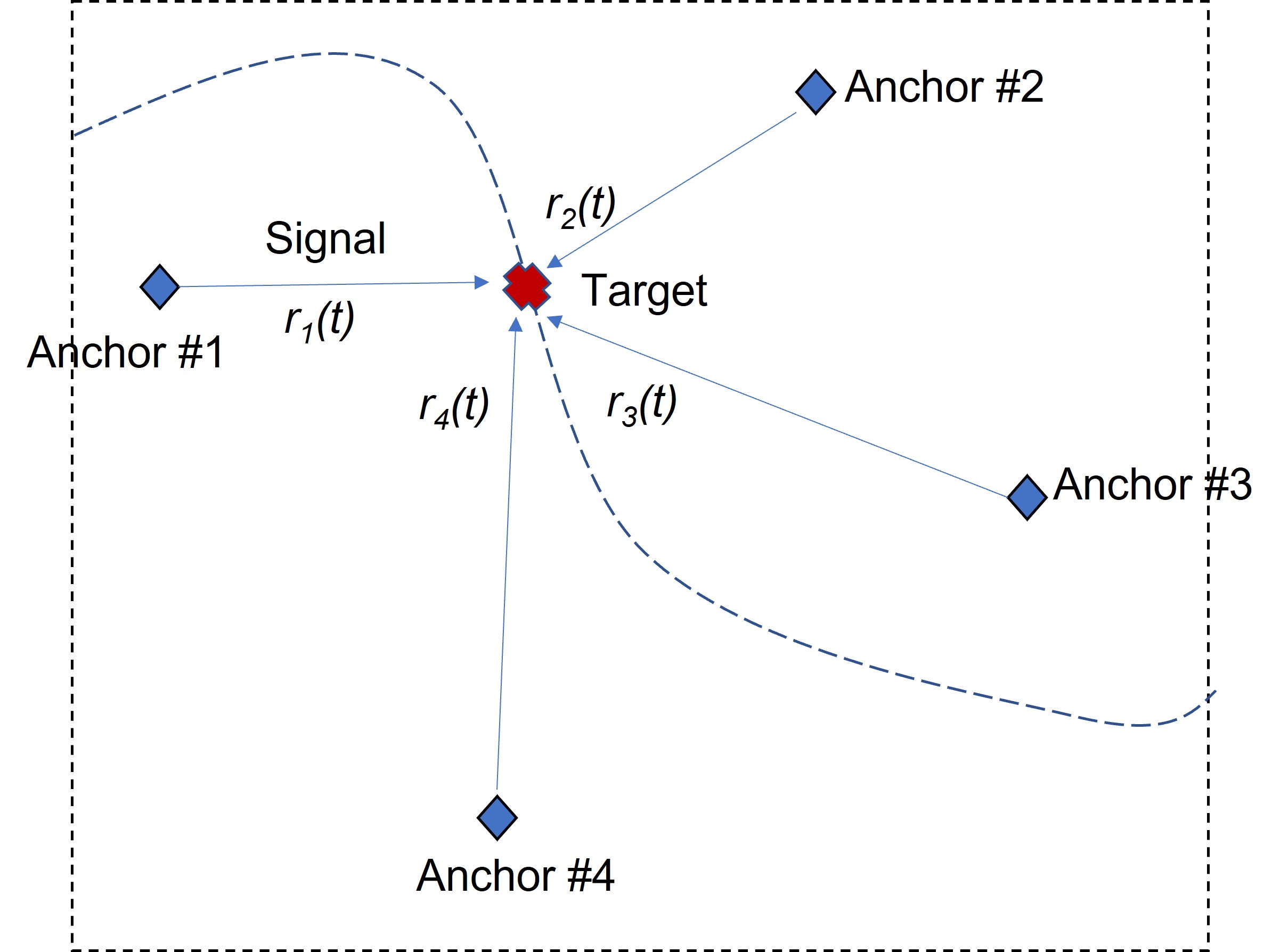}
	\vspace{3mm}
	\caption{System model: a set of anchors with the signals used for performing an active tracking. The dashed line is an example of trajectory in the considered space, delimited by the dashed square.}
	\label{fig:Fig_scenario}
\end{figure}

\section{System Model}
\label{Sec.:System Model}

The 2D network of wireless devices $\mathbf{N_{2D}}$, which is our reference layout, is inscribed in the general framework of localization and tracking in wireless networks, as in \cite{Moe.Z.Win2010,Moe.Z.Win2016,AnnaGuerra2018}.
Specifically, we consider a wireless network consisting of $N_{A}$ anchors and a target to be located. The target location $\mathbf{p} = \left[x, y \right]^{\mathrm{T}}$ is estimated based on the waveforms received from (downlink) or transmitted to (uplink) the anchors, as sketched in Fig. \ref{fig:Fig_scenario}. The set of anchors is denoted by $\underline{\mathrm{A}}$ = $\left\{1,2, \ldots, N_{A}\right\}$ and the position of the $j$-th anchor is assumed to be known and fixed and is denoted $\mathbf{p}_{j} = \left[x_{j}, y_{j}\right]^{\mathrm{T}}$ with $j \in \underline{\mathrm{A}}$. The work presented here can be extended to the 3D case in a straightforward way. 

For concreteness, let us consider the downlink. The signal $s_j(t)$ transmitted from the $j$-th anchor is a sequence of $M_j$ filtered symbols (impulses). The time spacing between two successive symbols, hereafter called $T_S$, is assumed to be the same for all the anchors. The energy of $s_j(t)$ is denoted $E_j$, so the energy per symbol is $E_j/M_j$. A standard result is that, in the small error regime and for pure Line-of-Sight (LoS) channels with Additive White Gaussian Noise (AWGN) with two-sided power spectral density $N_0/2$ equal for all the channels, the CRB below the mean square error achieved by any ToA-based unbiased estimator of the distance between the target and the $j$-th anchor is 
\begin{equation}\label{equ:CRB_ToA}
CRB_{dis,j} = \frac{c^2}{8 \pi^2 B^2 SNR_j},
\end{equation}
where $B$ is the effective bandwidth of $s_j(t)$ (which is hereafter assumed to be equal for all the anchors) and $SNR_j$ 
is the Signal-to-Noise ratio,
\begin{equation} \nonumber
SNR_j = \frac{E_j}{N_0}
\end{equation}
(see, for instance, \cite{Van_Trees_1968}). 

In case of multipath, the $j$-th waveform received by the target is
\begin{equation} \label{eq:r_k}
	r_{j}(t) = \sum_{l=1}^{L_{k}} \alpha_{j}^{(l)} s_j \left(t-\tau_{j}^{(l)} \right) + n_{j}(t), 
\end{equation}
where $\alpha_{j}^{(l)}$ and $\tau_{j}^{(l)}$ are the amplitude and delay, respectively, of the $l$-th path, $L_{j}$ is the number of multipath components, and $n_{j}(t)$ is the AWGN.

According to the very general and widely accepted model of \cite{Moe.Z.Win2010}, we assume that the system detects the LoS paths characterized by the set of parameters 
$\{\alpha_{j}^{(1)},\tau_{j}^{(1)}\}$ by means of appropriate procedure. Actually, in absence of a-priori knowledge of channel parameters, only these paths contribute to the detection of the position \cite{Moe.Z.Win2010}. If the link between one anchor and the target does not contain any LoS path, then we assume that this condition is detected by some appropriate procedure and that the anchor is muted.
With these assumptions, the relation between the target position and the delay of the LoS propagation path is 
\begin{equation} \label{eq:delay_1antenna}	 
\tau_{j}^{(1)}=\frac{d_j}{c},
\end{equation}
where $\tau_{j}^{(1)}$ is the delay of the LoS path, $c$ is the speed of the light and $d_j$ is the distance between the target and the $j$-th anchor:
\[d_j=\sqrt{(x-x_j)^2+(y-y_j)^2}.\] 
Given these premises, the analysis of \cite{Moe.Z.Win2010} shows that the CRB$_{dis,j}$ still has the same form as (\ref{equ:CRB_ToA}), but with    
\begin{align}\label{SNR_k}
    SNR_j & = \gamma_jE_j, 
\end{align}
where 
\begin{equation} \label{eq:gamma_j}	
\gamma_j = \frac{(1-\chi_j)|\alpha_j^{(1)}|^2}{N_0}.
\end{equation}
  The real number $\chi_j$, with \[1 >\chi_j \geq 0,\] is a multipath factor accounting for the average effect on the energy of the impulse associated to LoS path of the superposition of the impulse response of the band-limited transmit filter associated to the LoS path with the impulse responses associated to the Non LoS (NLoS) paths. Although in \cite{Moe.Z.Win2010} $\chi$ is kept separate  from the SNR, we prefer to incorporate it in the SNR to simplify the formulas. In a wideband system, where the impulse response of the transmit filter tends to the Dirac delta, $\chi$ tends to zero; besides \cite{Moe.Z.Win2010}, see also \cite{8269035} for the impact of  NLoS paths on the energy of the LoS one. In the numerical results hereafter presented, $\chi_j$ will be assumed to be equal for all the links, as they share similar propagation conditions, w/o loss of generality. As it is apparent from (\ref{SNR_k}), $\gamma_j$ is the key channel parameter, as it incorporates all the effects of the $j$-th channel on CRB$_{dis,j}$.

In the tracking procedure analyzed in the following sections, $\gamma_j$ is assumed to depend on the time-varying distance between the moving target and the $j$-th anchor and on the channel, while the energy $E_j$ is adjusted along time to optimize the trade of between accuracy of positioning and energy spent in the transmission.


%
%
%
%
%
%
%

\section{The recursive posterior Cramer-Rao bound}
\label{Sec:Recursive PCRB}

Among the very many papers that present theory and applications of the recursive PCRB, we found \cite{RobertM.TaylorJr2003} particularly appropriate to our scenario paper. 
Here, after revising the main steps, we introduce the relation between the predicted error covariance matrix of the position error and the variation of the energy of the signal transmitted from the anchors at the next tracking step, which  will be the basis for the material presented in the next section.

The derivation of the PCRB in the small error regime is based on the linearization of the measurement function, similarly to the common approach exploited for the EKF. Since the Kalman filter minimizes the mean square error, the approach is to impose the constraint on the accuracy of the estimated position on the error covariance matrix that is recursively computed at each step of the iteration of the EKF. 
Skipping the detailed presentation of the EKF, to make the paper readable, it is worth showing how our analysis and the simulation results to be presented in the final section are mapped inside this framework. The iteration for the state process is denoted as 
\begin{equation} \label{equ:state}
\mathbf{x}_{k} = \mathbf{F} \mathbf{x}_{k-1} + \mathbf{w}_{k},
\end{equation}
where $\mathbf{x}_{k}$ is the  state vector of size $N_x \times 1$ of the target at time $k$ (the subscript below vectors and matrices denotes the time index), the $N_x \times N_x$  matrix $\mathbf{F}$ is the process matrix that imports the model of target random motion, and the $N_x \times 1$ vector $\mathbf{w}_{k}$ is the state process noise vector, assumed to be a zero-mean Gaussian processes non-correlated in time with $N_x \times N_x$ covariance matrix $\mathbf{Q}_k$, generally time-dependent.  


In the computer simulations that we are going to present, $N_x=4$; the first two entries of the state vector are the two positions of the target, while the last two entries are the two velocities of the target, the latter two being modeled as random walks. The random steps of the two velocities are contained in the last two entries of $\mathbf{w}_{k}$, while the first two entries of $\mathbf{w}_{k}$ are zero. The $4 \times 4$ matrix $\mathbf{Q}_k$ is diagonal, with the first two elements of the main diagonal equal to zero and the last two elements equal to the square of the step size of the random velocity walk. Other common motion models are extended to $N_x = 6$ elements in 2D with the inclusion of the acceleration components. Then the measurement process is
\begin{equation} \label{equ:measurement}
\mathbf{z}_{k} = h(\mathbf{x}_{k}) + \mathbf{v}_{k},
\end{equation}
where $\mathbf{z}_{k}$ the $N_A \times 1$ measurement vector made by the $N_A$ distances between the target and the anchors and the $N_A \times 1$ measurement noise vector $\mathbf{v}_{k}$ is assumed to be a zero-mean Gaussian processes non-correlated in time and non-correlated with $\mathbf{w}_{k}$. Its  $N_A \times N_A$ covariance matrix $\mathbf{R}_{k}$ is assumed to be diagonal with elements equal to the $N_A$ $CRB_{dis,j}$ given in (\ref{equ:CRB_ToA}) at time $k$, the dependency on time being consequence of the variations of the SNR during the tracking process. The $N_A$-variate nonlinear function $h(\mathbf{x})$ calculates the $N_A$ distances from the actual position of the target to the $N_A$ anchors, thus importing in the measurement equation the known and fixed positions of the $N_A$ anchors.  

The $N_x \times N_x$ posterior covariance matrix $\mathbf{P}_{k|k}$ of the error between the actual state vector and the state vector estimated at time $k$ is obtained by the following predict and update steps:
\begin{equation}\label{eq:Pred_cov_matrix}
\mathbf{P}_{k|k-1} = \mathbf{F} \mathbf{P}_{k-1|k-1} \mathbf{F}^T + \mathbf{Q}_k, 
\end{equation}
\begin{equation}\label{eq:Cov_matrix}
\mathbf{P}_{k|k} = (\mathbf{I} - \mathbf{K}_{k} \mathbf{H}_k) \mathbf{P}_{k|k-1}, 
\end{equation}
where 
$\mathbf{P}_{k|k-1}$ is the prior error covariance matrix, the $N_x \times N_A$ gain matrix $\mathbf{K}_{k}$ is 
\begin{align}\label{eq:Kalman_gain}
\mathbf{K}_{k} & = \mathbf{P}_{k|k-1} \mathbf{H}_k^T (\mathbf{H}_k \mathbf{P}_{k|k-1} \mathbf{H}_k^T + \mathbf{R}_{k})^{-1} ,
\end{align} 
and
\[\mathbf{H}_k=\frac{\partial h(\mathbf{x})}{\partial \mathbf{x}} \bigg\rvert_{\hat{\mathbf{x}}_{k|k-1}} \]
is the $N_A \times N_x$  measurement matrix, that is the Jacobi matrix of the measurement function evaluated in the state $\hat{\mathbf{x}}_{k|k-1}$ predicted at time $k$. The $N_A \times N_A$ matrix  
\begin{equation}\label{equ:S}
\mathbf{S}_{k} = \mathbf{H}_k \mathbf{P}_{k|k-1} \mathbf{H}_k^T + \mathbf{R}_{k},
\end{equation} 
whose inverse is the rightmost term of \eqref{eq:Kalman_gain}, is the covariance matrix of the so-called \emph{innovations}, the errors between the measured distances and the predicted ones. 

The PCRB below the mean square error achievable by any unbiased estimator of target position at time $k$ is the sum of the first two elements of the main diagonal of the posterior error covariance matrix, which corresponds to the posterior errors on the two target coordinates:
\[\mbox{PCRB}_{pos,k}=P_{k|k}(1,1)+ P_{k|k}(2,2),\]
where $P_{k|k}(i,j)$ is the element $(i,j)$ of matrix $\mathbf{P}_{k|k}$.

\section{Constrained Energy Minimization}
\label{Sec.:Problem Formulation}

In the downlink, at each step of the tracking process, the total transmitted energy $E_{T}$ (for the sake of brevity, we skip the time index),
\begin{equation}\label{eq:energy_cost}
   \begin{array}{l}
   E_{T} = \sum_{j = 1}^{N_A} E_{j},
    \end{array}	
\end{equation}
is minimized with respect to the entries of the column vector $\mathbf{E}$ of the energies,
\[\mathbf{E}= (E_{1}, E_{2}, \cdots, E_{N_A})^T,\]
%
with the constraints
\begin{equation}\label{eq:constraints}
	\left\{\begin{array}{l}
				E_{j}M_j^{-1} \leq E_{MAX},\ \ j=1,2, \cdots, N_A, \\
    PCRB_{pos} = PCRB_{TARGET} 
	\end{array}\right.
\end{equation}
where $E_{MAX}$, which is assumed to be the same for all the transmitters, is the maximum energy per symbol that the transmitter can provide.
When we want to minimize latency in a system where anchors transmit in time division multiplexing, the energies per symbol are set to $E_{MAX}$ and the optimization acts only  the set $\{M_j\}$. When desired, the $j$-th anchor can be muted by putting $M_{j}=0$. 

In the uplink, the localization is performed by the anchors, or a device in the network connected to the anchors, by using the signals that are received from the target (Fig. \ref{fig:Fig_scenario}).
In this scenario, the total energy transmitted by the single target can be adapted in order to perform the localization effectively. 
The formal representation of the cost function is equivalent to the downlink case, but with only one transmitter, i.e. the only allowed value of $j$ is $j=1$. 

Finally, as it will be remarked also in the algorithms, we observe that the minimization of the energy can be translated into the modification of two physical parameters at the transmitter side: the energy per symbol and/or the number of symbols that compose the transmitted signal.

\begin{figure}[tb!]
	\centering
 	\includegraphics[scale=0.5]{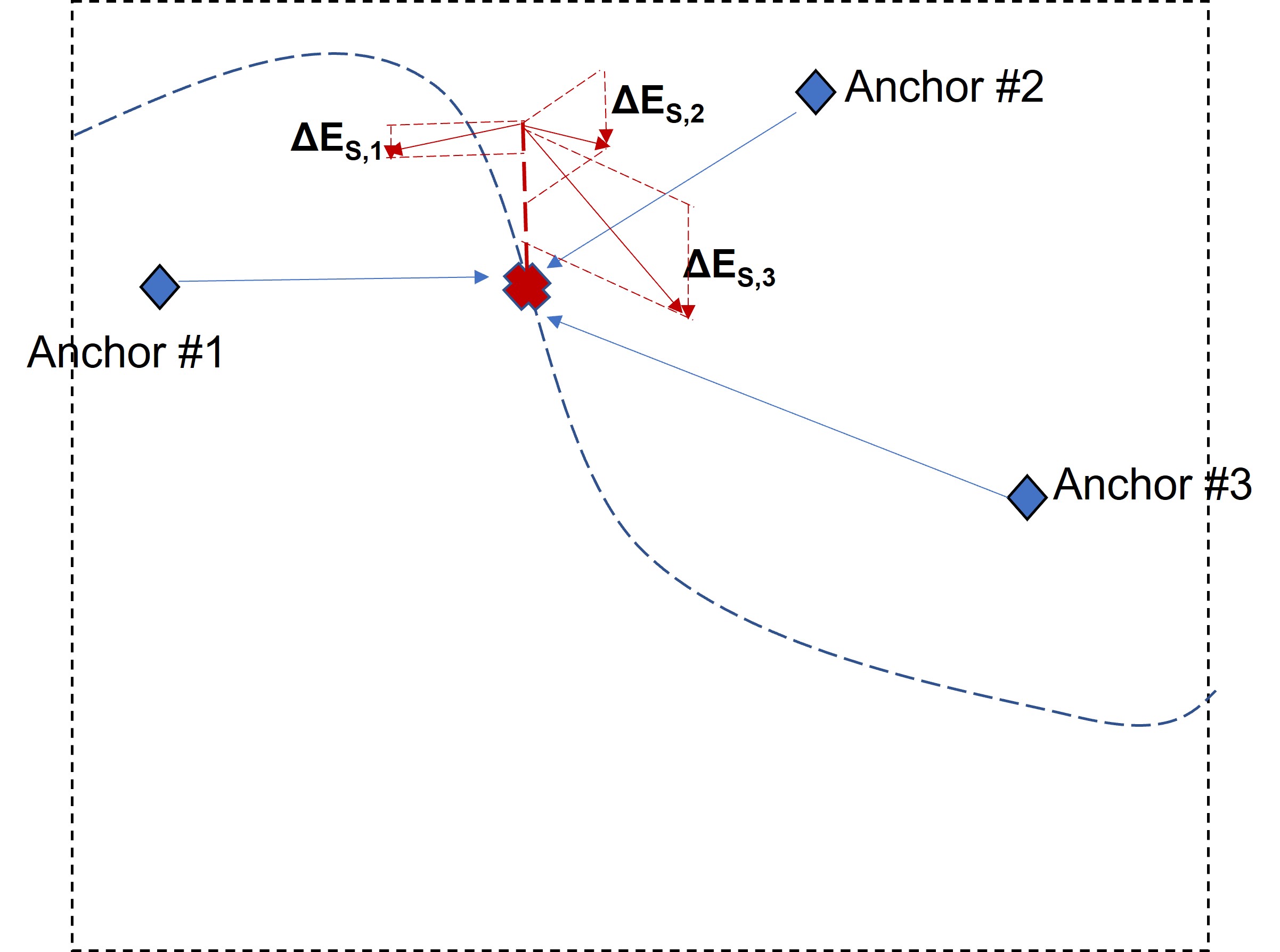}
 	\vspace{3mm}
 	\caption{Representation of the algorithm for minimizing the energy during a trajectory tracking.}
 	\label{fig:DA_scenario}
\end{figure}

\subsection{First order variation of the PCRB}
\label{subsec:First_order_variation_PCRB}

The solution of the optimization problem would require that, at each step of the tracking procedure, the PCRB is expressed as a function of the transmission energies. We denote this function as    
\begin{equation*}
PCRB_{pos,k}(\mathbf{E}) = f_k(\mathbf{E}),
\end{equation*}
where the subscript $k$ imports all the history of tracking till time $k$ included, and it is understood that the PCRB given in the previous section is obtained when $\mathbf{E}=\mathbf{E}_k$, that is
$f_{k}(\mathbf{E}_k)=PCRB_{pos,k}$.
Then, the constrained optimization of the energy at time $k$ requires that the entire function $f_k(\mathbf{E})$ is included in the KKT conditions together with the constraints
(\ref{eq:constraints}).
The extremely high complexity of this approach can be substantially reduced by exploiting the following analysis of the first order variations.

Let us consider the first order variations of the PCRB around the current vector of energies $\mathbf{E}_k$:
\begin{equation}\label{equ:pcrbvsenergy}
f_k(\mathbf{E}) -f_k(\mathbf{E}_k)\approx  (\mathbf{E}^T-\mathbf{E}_k^T) \nabla f_k(\mathbf{E})|_{\mathbf{E}_k},
\end{equation}
where $\nabla$ is the gradient with respect to the $N_A$ energies.  
In App. \ref{App:PCRB_gradient} it is shown that 
\begin{equation}\label{equ:mainresult}
\nabla f_k(\mathbf{E})|_{\mathbf{E}_k}=(q_{1,k},q_{2,k}, \cdots, q_{N_A,k})^T,
\end{equation}
with
\begin{equation} \label{equ:Delta_P_qj}
q_{j,k} = - \frac{(K_k(1,j)^2 + K_k(2,j)^2) CRB_{dis,j,k}}{E_{j,k}},
\end{equation}
and, 
\begin{equation}\label{equ:discrete_pcrbvsenergy}
f_k(\mathbf{E}) -f_k(\mathbf{E}_k) \approx  \sum_{j = 1}^{N_A} q_j \cdot (E_j-E_{j,k}).
\end{equation}


In \eqref{equ:Delta_P_qj} we have highlighted that $CRB_{dis,j,k}$ depends on the time $k$ as the received energy depends on the transmitted energy, which is the objective of the constrained minimization, and also on the instantaneous channel realization.   
We observe that the variation $\mathbf{E}^T-\mathbf{E}_k^T$ produces a variation of the covariance $\mathbf{R}_{k}$ since the $j$-th ToA measure quality depends on the signal energy associated to the corresponding link target - anchor $j$ ($j = 1, \cdots, N_A$), as explained in Sect. \ref{Sec.:System Model}. 
Therefore, according to \eqref{SNR_k} and \eqref{equ:CRB_ToA}, a variation $\Delta E_{j,k}$ in the transmitted energy generates a first order variation 
\begin{align}\label{equ:Delta_CRB_ToA}
\Delta CRB_{dis,j,k} &= CRB_{dis,j,k}(E_{j,k} + \Delta E_{j,k}) + \\
& \quad - CRB_{dis,j,k}(E_{j,k}) \\
& \approx - CRB_{dis,j,k} \frac{\Delta E_{j,k}}{E_{j,k}},
\end{align}
in the CRB of the j-th link after applying the energy variation $\Delta E_{j,k}$, and in the PCRB, by means of \eqref{equ:Delta_P_qj} and \eqref{equ:discrete_pcrbvsenergy}. 
%
%
%
%
%
%
%
%
%
%
In fact we can estimate the impact of energy variations on the PCRB rewriting \eqref{equ:discrete_pcrbvsenergy} as
\begin{equation}\label{equ:Delta_P_final}
\Delta PCRB_{pos,k} \approx \sum_{j = 1}^{N_A} q_{j,k} \cdot \Delta E_{j,k}.
\end{equation} 
%
%
%
From the perspective of the application to algorithms for the energy optimization, the parameters $q_{j,k}$ allow to express in closed form the first-order variations of the tracking error as a function of the transmitted energies $E_{j,k}$ according to the channel conditions imported in the CRB by the channel parameters $\gamma_j$ in \eqref{eq:gamma_j}. We observe that (i) each coefficient $q_{j,k}$ is negative as an increase of the energy corresponds to a decrease of the PCRB and vice-versa and (ii) increasing the channel gain of the j-th link, or $\gamma_j$, the absolute value of $q_{j,k}$ decreases, meaning that the impact of an energy variation in that link has less impact on the overall PCRB.
When, in the uplink, we have only one signal transmitted by the target and received by all the anchors, the relation \eqref{equ:Delta_P_final} becomes simply
\begin{equation}\label{equ:Delta_P_final_uplink}
\Delta PCRB_{pos,k} = \Delta E_{k} \sum_{j = 1}^{N_A} q_{j,k}.
\end{equation} 
%


\subsection{Energy optimization based on the PCRB (JTE)}
\label{subsec:delta_algorithm}
\begin{figure}[t]
	\centering
 	\includegraphics[width=0.6\textwidth]{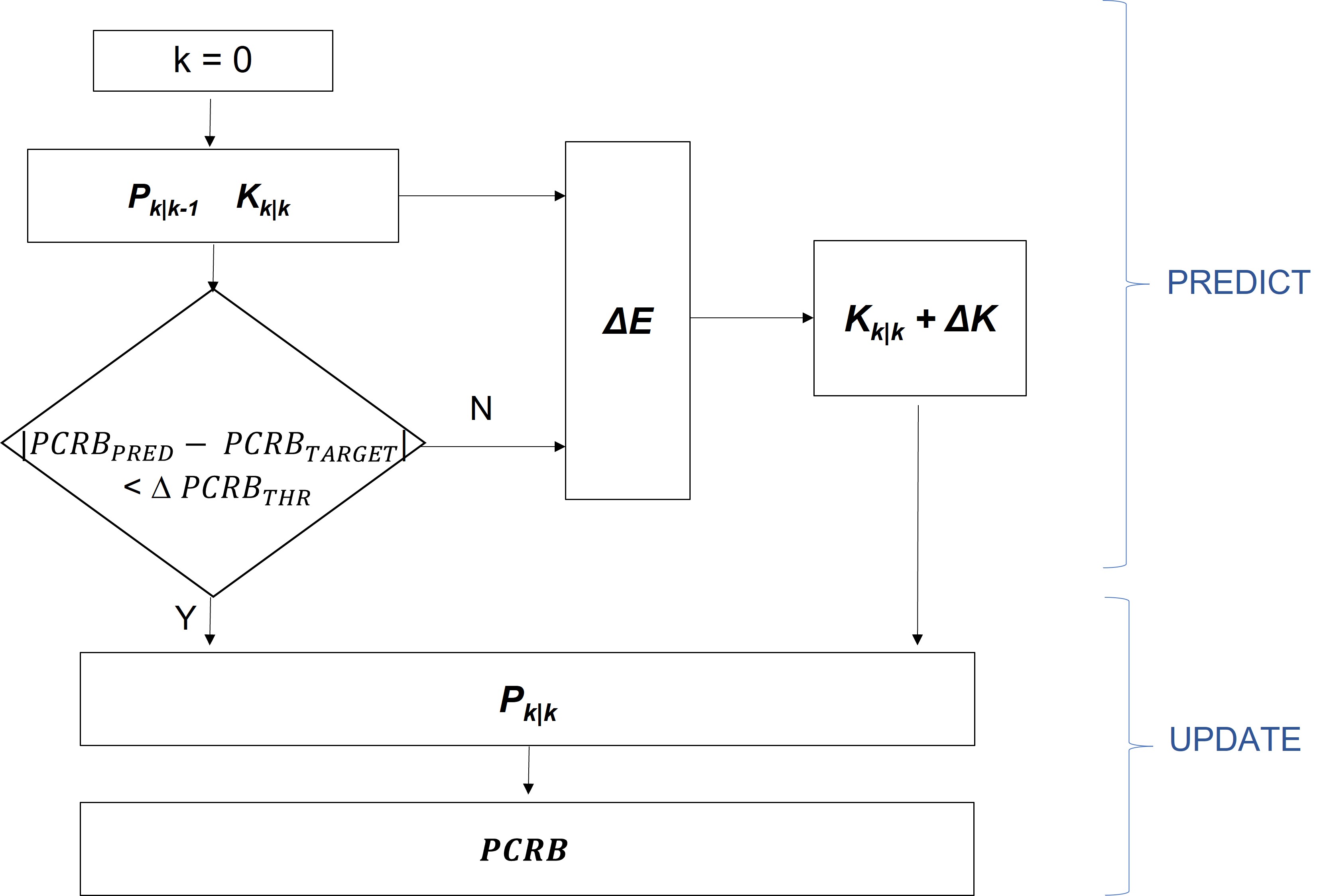}
 	\vspace{3mm}
 	\caption{Flowchart of the JTE algorithm for the solution of the downlink/uplink latency minimization.}
 	\label{fig:delta_algorithm}
\end{figure}

This sub-optimal method exploits the first order analysis of the PCRB, provided in \eqref{equ:Delta_P_final}, i.e. the impact of energy variations in the anchors signals on the PCRB according to the current tracking state and it can be considered an example of a \textit{constrained gradient descent method}, where the direction of the cost function minimization \eqref{eq:energy_cost} is determined according to its impact on the PCRB. We consider the more general case of downlink transmission with $N_A$ different signals from the anchors, as the uplink case is a straightforward modification, making use of \eqref{equ:Delta_P_final_uplink}.
The flow diagram of the algorithm, denoted as Joint Tracking and Energy variation algorithm (JTE) is reported in Fig. \ref{fig:delta_algorithm}: at each predict step of the PCRB iterations, it is evaluated the possibility of changing the energy allocation while respecting the performance constraint. The possbile variation of the energy allocation is incorporated in the update phase of the PCRB iteration, updating the gain and covariance matrices. The algorithm is detailed in the following points.

\textbf{Initialization}:
 \begin{itemize}
    \item[I1)] $\Delta E_{MAX} > 0$ is the maximum energy variation allowed at each step.
    \item[I2)] $E_{MAX}$ and $E_{MIN}$ are maximum and minimum energies that are allowed at each link. If $E_{MIN} = 0$, a generic link can be excluded from the tracking process if its allocated energy goes below zero, also temporarily.
    \item[I3)] A threshold $\Delta PCRB_{THR}$ is defined in order to activate the energy modification only when the predicted advantage on the PCRB is higher than $\Delta PCRB_{THR}$. 
 \end{itemize}

\textbf{Description of the algorithm}:
 \begin{itemize}
  \item[1)] At each step $k$, after the predict phase of the PCRB iterative estimation, the covariance matrix $\mathbf{P}_{k|k-1}$ and the Kalman gain $\mathbf{K}_{k}$ are computed. The predicted positioning error is computed as
  \begin{equation}\label{eq:PCRB_PRED}
 	PCRB_{PRED,k} = \mathbf{P}_{k|k-1}(1,1) + \mathbf{P}_{k|k-1}(2,2).
  \end{equation}
  \item[2)] If the difference $\Delta PCRB = |PCRB_{PRED,k} - PCRB_{TARGET}|$ is lower than $\Delta PCRB_{THR}$ there is no variation of the energy and go to step 4. If $\Delta PCRB_{THR} = 0$ the energy variation is always analyzed.   
  \item[3)] The values of the positive coefficients $|q_{i,k}|$ in \eqref{equ:Delta_P_final} are computed and ordered in ascend order in the vector $\mathbf{q}$. The corresponding ordered list of anchors is $\mathbf{L}_A$, from the lowest $|q_{i,k}|$ to the highest.
  \item[4)] The vector of variations $\mathbf{\Delta E}$ is computed according to the following iterations, numbered from $i = 1$ to $N_A$:
 	\begin{itemize}
 	\item[(a)] $i=1$, $\Delta PCRB_i = \Delta PCRB$.
 	\item[(b)] If $PCRB_{PRED,k} < PCRB_{TARGET}$, i.e. the spent energy can potentially decrease, take the anchor $n$ corresponding to the index $i$ of $\mathbf{L}_A$ and compute the $\Delta E_n = - min \{ \Delta E_{MAX}, \Delta PCRB_i/|q_{i,k}| \}$. If $E_n + \Delta E_n < E_{MIN}$, correct $\Delta E_n$ in order to respect the minimum $E_{MIN}$. Then update $\Delta PCRB_{i+1} = \Delta PCRB_i - \Delta E_n \cdot |q_{n,k}|$.
	\item[(c)] If $PCRB_{PRED,k} > PCRB_{TARGET}$, i.e. the spent energy should increase, take the anchor $n$ corresponding to the index $N_A - i + 1$ of $\mathbf{L}_A$ and compute the $\Delta E_n = min \{ \Delta E_{MAX}, \Delta PCRB_i/|q_{i,k}| \}$. If $E_n + \Delta E_n > E_{MAX}$, correct $\Delta E_n$ in order to respect the maximum $E_{MAX}$. Then update $\Delta PCRB_{i+1} = \Delta PCRB_i + \Delta E_n \cdot |q_{n,k}|$.
 	\item[(d)] If $i = N_A$ or $\Delta PCRB_i < \Delta PCRB_{THR}$ stop, otherwise $i=i+1$, move to the next anchor and go to step (b).
   \end{itemize}  
  \item[5)] If the energy allocation has been changed, update the Kalman gain $\mathbf{K}_{k}$ by means of
  \begin{equation*}
     \mathbf{\Delta K} \approx - \mathbf{P}_{k|k-1} \cdot \mathbf{H}^T_k \cdot \mathbf{S_k}^{-1} \cdot \mathbf{\Delta S} \cdot \mathbf{S_k}^{-1},
  \end{equation*}
  where $\mathbf{\Delta S}$ is computed with \eqref{equ_app:dS_dE} in App. \ref{App:PCRB_gradient}.
  \item[6)] Proceed with the update step of PCRB, computing $\mathbf{P}_{k|k}$.  
  \item[7)] $k = k +1$ and proceed to the next step.
 \end{itemize}

The rationale behind the ordered steps on the vector $\mathbf{q}$ is that the reduction of the energy is done according to the objective of minimizing the total energy limiting the impact on the $PCRB$ as much as possible: more energy is subtracted from the anchors with lowest $q_i$ when the energy can be reduced and vice-versa.  
The complexity of this algorithm does not increase the complexity of the original PCRB iterative process and, correspondingly, also of a possible real implementation, based on the EKF-based tracking. 

An additional remark is necessary in the implementation of step (b). When the energy is changed, two physical parameters at the transmitter side could be modified: the power, until it remains below a maximum level, and/or the number of symbols in the transmitted sequence.

\begin{figure}[tb!]
	\centering
 	\includegraphics[scale=0.5]{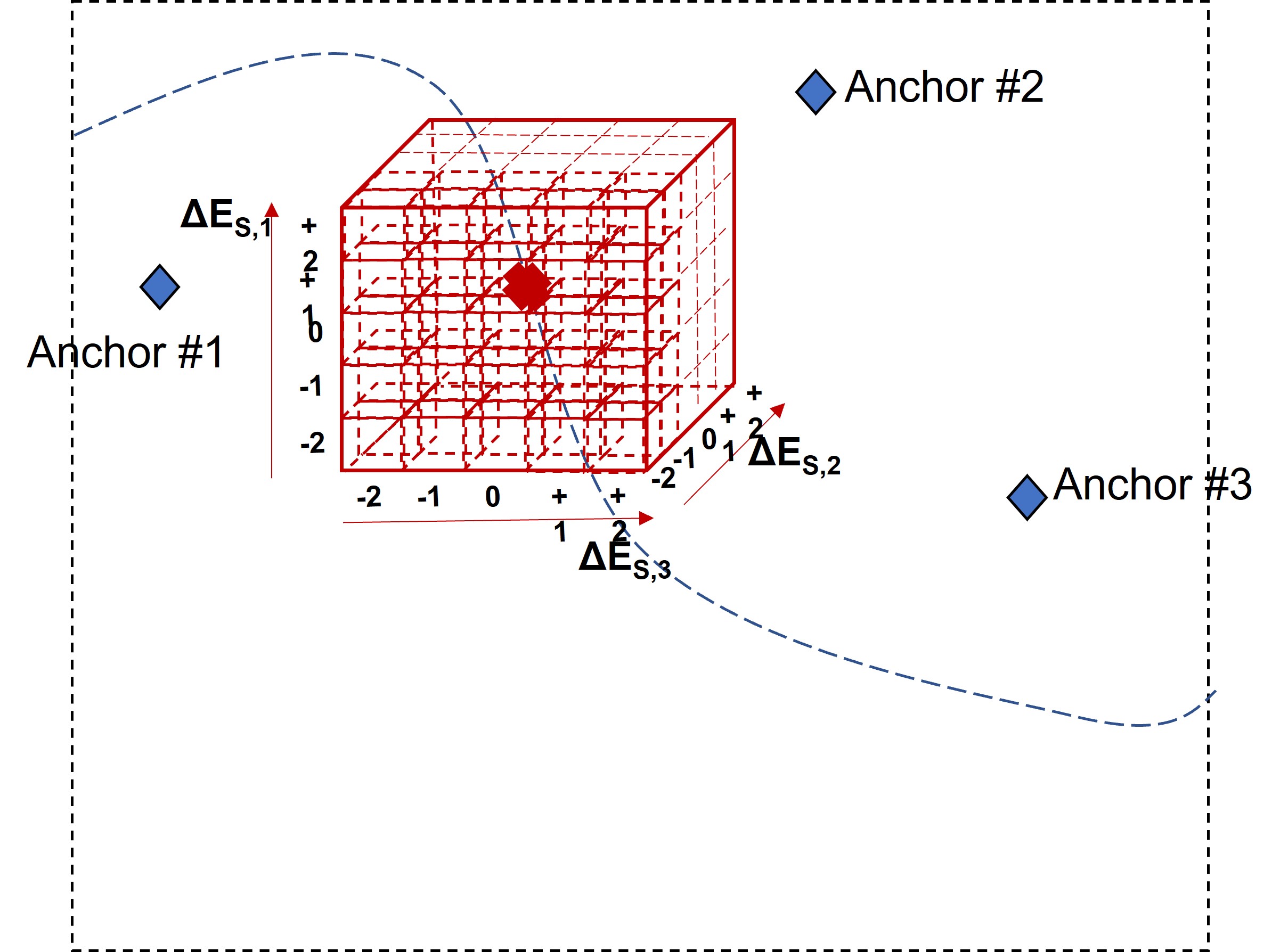}
 	\vspace{3mm}
 	\caption{Representation of the proposed semi-exhaustive search during trajectory tracking. Only three anchors are sketched, just for drawing the three-dimensional search space.}
 	\label{fig:SSA_scenario}
\end{figure}

\subsection{Energy optimization based on semi-exhaustive search (SSW)}
\label{subsec:sliding_space}
In order to perform the minimization of the cost function \eqref{eq:energy_cost}, here we follow a step-by-step process, in which around the current tracking state an exhaustive search of all the combinations of energy variations is performed for selecting the one characterized by the minimum overall energy within the performance requirements. The computational complexity of this approach is reduced by limiting the search space of the combinations of energy variations $\Delta E_j$ (w.r.t. the current ones at each step $k$) to a subspace defined by the maximum energy variation that can be imposed per anchor and a discrete energy variation step ($E_{SSW,STEP}$, for computing the set of discrete combinations). In Fig. \ref{fig:SSA_scenario}, this search is sketched for three anchors and a search space delimited by testing all the $\Delta E_j$ between $-2 E_{SSW,STEP}$ and $+2 E_{SSW,STEP}$ for all the anchors ($j = 1, \cdots,N_A = 3$).  

The detailed description of the process, denoted as Semi-exhaustive Search over a Window of energy variations (SSW), and its integration in the iterative PCRB computation is reported in the following steps.

\textbf{Initialization}:
\begin{itemize}
 \item[I1)] Define the energy variation step $E_{SSW,STEP}$ as a sub-multiple of the maximum energy variation that can be imposed at each anchor per step, i.e. $\Delta E_{MAX} = m \cdot E_{SSW,STEP}$. 
 \item[I2)] The energy variation at each step per anchor is discretized: the energy that can be added to or subtracted from the $j$-th anchor is the $j$-th element of the $(N_A \times 1)$ vector $V_{SSW} = E_{SSW,STEP} \cdot [-m, -(m-1), \cdots, 0, \cdots, (m-1), m ]^T$.
 \item[I3)] According to the possible $(2m+1)$ elements in $V_{SSW}$ per anchor, the $(N_{SSW,COMB} \times N_A)$ list $L_{SSW,COMB}$ of all the possible combinations of $N_A$ elements, one for each anchor, is generated; the number of combinations $N_{SSW,COMB}$ is equal to $(2m+1)^{N_A}$.   
\end{itemize}
 
\textbf{Description of the algorithm}:
\begin{itemize}
 \item[1)] At each step $k$ (as in JTE, \ref{subsec:delta_algorithm}), after the predict phase of the PCRB iterative estimation, the covariance matrix $\mathbf{P}_{k|k-1}$ and the Kalman gain $\mathbf{K}_{k}$ are computed. Then the predicted positioning error $PCRB_{PRED,k}$ is computed as in \eqref{eq:PCRB_PRED}.
 \item[2)] The values of the coefficients $q_{i,k}$ in \eqref{equ:Delta_P_final} are computed.
 \item[3)] For each combination $\Delta E_i$ in $L_{SSW,COMB}$ $(i = 1, 2, \cdots, N_{SSW,COMB})$, compute the PCRB variation $\Delta PCRB_i$ according to \eqref{equ:Delta_P_final} and $PCRB_{i} = PCRB_{PRED,k} + \Delta PCRB_i$. The $i$-th combination that (i) respect the performance requirement $PCRB_{i} \leq PCRB_{TARGET}$ and (ii) is characterized by the minimum total energy variation $min_i \{\sum_{n=1}^{N_A} \Delta E_i(n)\}$ (possibly negative if the system is able to save energy, otherwise positive) is selected for updating the current vector of energies: $\mathbf{E}_{k} = \mathbf{E}_k + \mathbf{\Delta E_{i,k}}$ and the corresponding $\mathbf{K}_k + \mathbf{\Delta K}$, $\mathbf{P}_{k|k}$ are used for the PCRB update step (as in steps 5 and 6 of JTE).
 \item[4)] $k = k+1$ and proceed to the next step.       
\end{itemize}

Finally, as stated for the previous JTE algorithm, the energy variations can be translated, for their physical interpretation, in power and/or number of transmitted symbols.

\subsection{Energy optimization benchmark based on semi-exhaustive search (SSW-benchmark)}
\label{subsec:sliding_space_benchmark}
The same semi-exhaustive search described in \ref{subsec:sliding_space} is performed here with the exact computation of the PCRB for each combination of energy variations, instead of using the first order variation of the PCRB presented in Sect. \ref{subsec:First_order_variation_PCRB}. This version of the search is used either for validating the accuracy of \eqref{equ:Delta_P_qj} and \eqref{equ:pcrbvsenergy} or showing the computational advantage provided by the first order approximation.
W.r.t. SSW in \ref{subsec:sliding_space}, steps 2-3 are replaced by the following one, not using the first order approximation presented in sec. \ref{subsec:First_order_variation_PCRB}:
\begin{itemize}
\item[2] For each combination $\Delta E_i$ in $L_{SSW,COMB}$ $(i = 1, 2, \cdots, N_{SSW,COMB})$, compute the new total energy $\mathbf{E}_{i,k} = \mathbf{E}_k + \mathbf{\Delta E_i}$, update $\mathbf{R}_k$ to $\mathbf{R}_{i,k}$ accordingly and perform the update step of the PCRB, for computing the exact $PCRB_{i}$. The operations to be performed are: 
    \begin{itemize}
     \item[(a)] $\mathbf{S}_{i,k} = \mathbf{H}_k \cdot \mathbf{P}_{k|k-1} \cdot \mathbf{H}_k^T + \mathbf{R}_{i,k}$;
     \item[(b)] $\mathbf{K}_{i,k} = \mathbf{P}_{k|k-1} \cdot \mathbf{H}_k^T \cdot \mathbf{S}_{i,k}^{-1}$;
     \item[(c)] $\mathbf{P}_{i,k|k} = (\mathbf{I} - \mathbf{K}_{i,k} \cdot \mathbf{H}_k) \cdot \mathbf{P}_{k|k-1}$;
     \item[(d)] $PCRB_i = \mathbf{P}_{i,k|k}(1,1) + \mathbf{P}_{i,k|k}(2,2)$.
    \end{itemize} 
The $i$-th combination that (i) respect the performance requirement $PCRB_{i} \leq PCRB_{TARGET}$ and (ii) is characterized by the minimum total energy variation $min_i \{\sum_{n=1}^{N_A} \Delta E_i(n)\}$ is selected for updating the current vector of energies, with the corresponding $\mathbf{K}_k = \mathbf{K}_{i,k}$, $\mathbf{P}_{k|k} = \mathbf{P}_{i,k|k}$, and $PCRB = PCRB_i$ (update step already computed for each combination).
\end{itemize}

\subsection{Complexity analysis}
\label{subsec:complexity}

The proposed approach for optimizing the energy consumption given a PCRB constraint has also some clear advantages from a computational point of view w.r.t. greedy strategies. Here we have considered the number of multiplications involved in the three algorithms described in subsections \ref{subsec:delta_algorithm} - \ref{subsec:sliding_space_benchmark}, i.e. JTE, SSW and SSW-benchmark.
It is clear that the complexity of the algorithms JTE and SSW is substantially reduced compared to the benchmark as all the matrix computations are avoided and, thanks to the first order approximation, it is possible to predict with few operations among scalars the impact on PCRB of the modified energies. On the other hand, the SSW-benchmark involves, at each step, a new matrix inversion and three new matrix multiplications for computing the PCRB for each possible combination. Moreover, the SSW and SSW-benchmark options are greatly affected by the size of the search space $N_{SSW,COMB}$ in which the combinations of energy variations are tested and selected. 

\begin{table}[bt]
    \centering
    \begin{tabular}{|c|c|c|}
        \hline 
       Algorithm  & Operations & Complexity \\
                  &            & (multiplications) \\
        \hline 
        \hline
       JTE & $N_A$ iterations  & $N_A$ \\
           & on $\Delta PCRB$   &  \\
        \hline
       SSW & $\Delta PCRB$ for  & $N_{SSW,COMB} \cdot N_A$ \\
           &  each combination &                          \\
        \hline
       SSW & $\mathbf{S}_k^{-1}$ & $N_{SSW,COMB} \cdot N_A^3 / 3$ \\
       benchmark   & $\mathbf{P}_{k|k}$ & $ N_{SSW,COMB} \cdot$ \\
                     & for each combination &  $(N_x^3 + N_x^2 N_A + N_x N_A^2)$ \\
       \hline
    \end{tabular}
    \caption{Computational complexity of the algorithms, measured by the number of multiplications between numbers at each step $k$.}
    \label{tab:computational_complexity}
\end{table}

\subsection{Interpretation in terms of latency optimization}\label{subsec:alg_latency_power}
In Sect. \ref{Sec.:System Model}, we have already presented the simple, proportional relation between the SNR and the number of transmitted symbol $M$ in \eqref{SNR_k}. This means that the optimization and the methods presented in the previous Sect. \ref{Sec.:Problem Formulation} are easily applicable also to a latency optimization, when the following scenario and assumptions are considered:
\begin{itemize}
	\item the transmission is downlink, from the anchors to the target.
	\item The target performs the signal acquisition and analysis according to a Time Division Multiplexing (TDM) of the signals, i.e. serially. Therefore, the transmission time of the signals from anchors will be equal to the sum of all the anchors transmission times. This assumption is coherent with the common, practical implementation of these systems, in which the target is not equipped with multiple receive chain that can operate in parallel.
	\item The target and the anchors can adapt the number of transmitted symbols in order to minimize the overall transmission time.
	\item The latency of the process is approximated by the transmission time or the transmission time is the only term that can be adapted in the overall latency budget.
\end{itemize}
In this scenario, the total transmission time $T_{LAT}$ is the sum of all the anchors transmission times $T_{j}$ and, also, it is proportional to the overall energy,
\begin{equation*}
 	T_{LAT} = \sum_{j = 1}^{N_A} T_{j} = T_S \cdot \sum_{j = 1}^{N_A} M_{j} = \frac{T_S}{E} \cdot \sum_{j = 1}^{N_A} E_{j}.
\end{equation*}
The same methods described in subsections \ref{subsec:delta_algorithm} and \ref{subsec:sliding_space} can be applied here, with some minor modifications, since the optimization is done w.r.t. number of transmitted symbols by each anchor $M_j$, which is a discrete, integer variable.

\begin{figure}[!tb]
	\centering
	\includegraphics[scale=0.68]{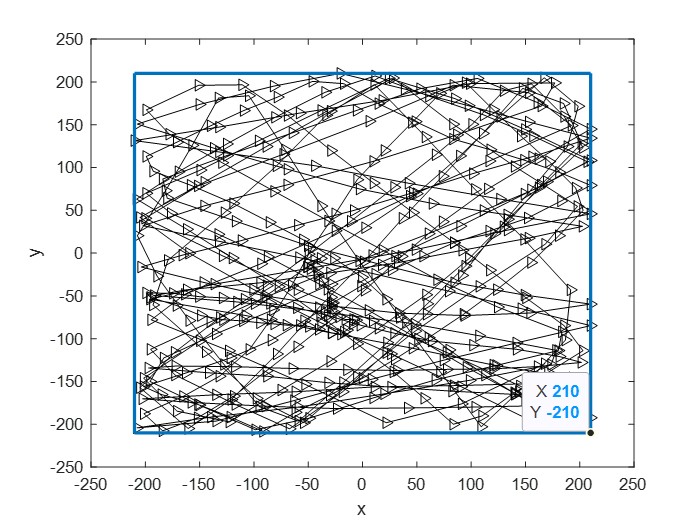}
	\vspace{3mm}
	\caption{Example of a trajectory with the adopted motion model within the target area.}
	\label{fig:Fig_traj_example}
\end{figure}

\begin{table}[bt]
\caption{The parameters of the optimization algorithms}
\label{tab:parameters}
\centering
\begin{tabular}{|c|c|c|}
    \hline
    Algorithm & \multicolumn{1}{|c|}{Parameter} & \multicolumn{1}{|c|}{Value}  \\
    \hline
    \hline
    All & $T_{EST}$ & $1.0$ s \\
    \hline
    All & $T_{OBS}$ & $532 \cdot T_{EST}$ s \\
    \hline
    \hline
    All & $PCRB_{TARGET}$ & $1$ $m^2$ \\
    \hline
    All & $E_{INIT}$ & $32 \cdot E$ (no multipath) \\
        &            & $52 \cdot E$ (multipath) \\
    \hline
    All & $E_{MIN}$ & $0$ \\
    \hline
    All & $E_{MAX}$ & $8 \cdot E_{INIT}$ \\
    \hline
    All & $E/N_0$	 & $0$ dB at $420$ m \\
    \hline
    JTE & $\Delta E_{MAX}$ & $5 \cdot E$ \\
    \hline
    JTE & $\Delta PCRB_{THR}$ & $0.05 \cdot PCRB_{TARGET}$ \\      
    \hline
    SSW & $E_{SSW,STEP}$ & $5 \cdot E$ \\      
    \hline
    SSW & $m = $ & $\{10,8,6\}$ \\      
        &      $\Delta E_{MAX}/E_{SSW,STEP}$  & for $N_A=\{4,5,6\}$ \\      
    \hline
    \hline
\end{tabular}
\end{table}

\begin{figure}[!b]
   \centering
   \subfloat[$\sigma_w = 0.025$]{
        \includegraphics[width=.48\textwidth]{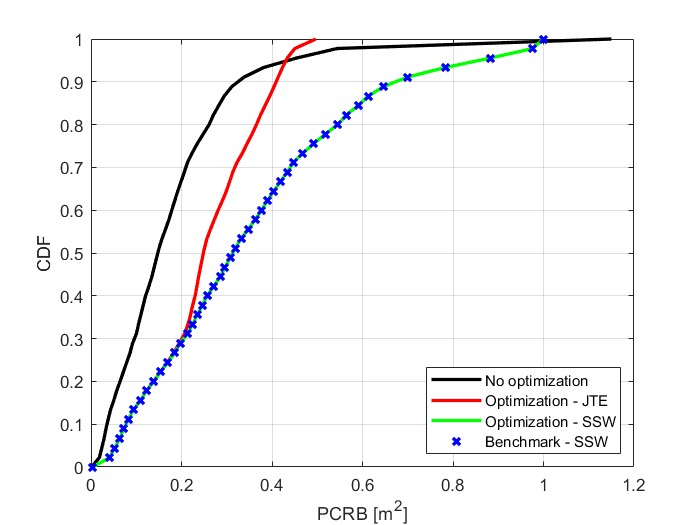}
        \label{Fig_CDF_PCRB:A}
    } 
    \subfloat[$\sigma_w = 1.0$]{
        \includegraphics[width=.48\textwidth]{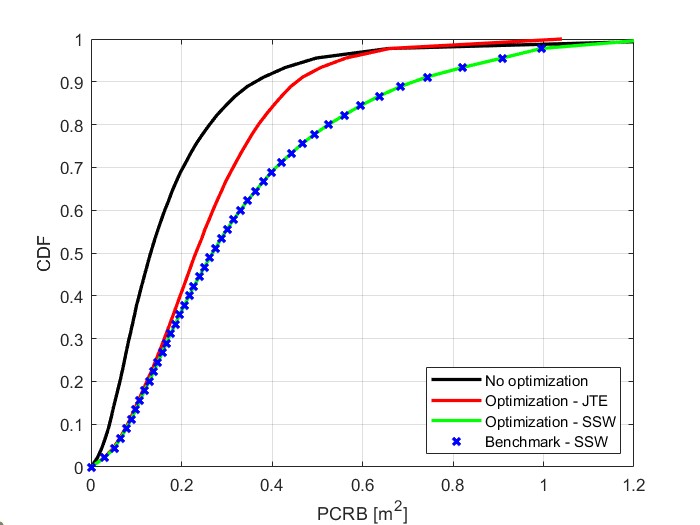}
        \label{Fig_CDF_PCRB:B}
    }
 	\caption{CDF of the PCRB during the trajectories for $\sigma_w = 0.025$ (a) and $\sigma_w = 1.0$ (b) with $N_A = 4$ anchors.}
    \label{fig:Fig_CDF_PCRB}
\end{figure}

\section{Numerical Results}\label{sec:numerical_results}

The numerical results are obtained in a scenario analogous to Fig. \ref{fig:Fig_scenario} and characterized by the following parameters:
\begin{itemize}
\item the side of the 2D square area is $420$ meters with $N_A = \{ 4,5,6 \}$ anchors randomly placed in the area.
\item Unless otherwise stated, the propagation is subject to a pathloss given by $\alpha_{PL} + \beta_{PL} \cdot log_{10}(d/d_0)$ with $\alpha_{PL} = 40$ dB, $\beta_{PL} = 30$ dB, $d_0 = 1$ m. 
\item The simulations are also performed with the additional presence of a Rice multipath model with Rice factor equal to $0$ dB, in order to check the impact of challenging multipath conditions (in which the direct LoS path power becomes comparable to reflections power, causing wide SNR fluctuations). 
\item The signal has a bandwidth $B = 10$ MHz and the mean SNR 
is equal to $0$ dB for a single symbol at a distance equal to the square side.  
\item The motion model is the random walk on the velocity (a discrete white noise acceleration model \cite{Shalom_2002}), in which the velocity at each step $k$ is updated in its two components by
\begin{align*}
v_{k,x} & = v_{k-1,x} + \sigma_w \cdot \delta_{k,x} \\
v_{k,y} & = v_{k-1,y} + \sigma_w \cdot \delta_{k,y} \\
\end{align*}
where $\sigma_w$ is the standard deviation of the random process, $\delta_{k,x}$, $\delta_{k,y}$ two normal random variables with zero mean and unitary variance and $v_{0,x} = v_{0,y} = 0$. The initial coordinates are the center of the area and each time the target reaches a boundary of the area, it is reflected till to the end of the trajectory (in Fig. \ref{fig:Fig_traj_example} there is an example). A number of $50$ trajectories is extracted and simulated for an entire observation time $T_{OBS}$. These choices allow the achievement of a virtually uniform distribution of the points crossed by the target in the area (the total number of simulated points in the area is approximately equal to $25600$).
\item The PCRB iterations are based on the tracking update by means of the reception of $N_A$ ToA measures with a period $T_{EST}$. With a $4 \times 1$ state vector $\mathbf{x}_{k} = [x_{k}, y_{k}, v_{k,x}, v_{k,y}]^T$ containing the current target position and velocity, the state and measurement matrices in \eqref{equ:state} and \eqref{equ:measurement}, according to the exemplary motion model, are given by 
\begin{equation*}
\mathbf{F} =   
	\left[ {\begin{array}{cccc}
    	  1 & 0 & T_{EST} & 0 \\
    	  0 & 1 & 0 & T_{EST} \\
    	  0 & 0 & 1 & 0 \\
    	  0 & 0 & 0 & 1 \\
  	\end{array} } \right]
\end{equation*}
with size $(4 \times 4)$, and
\begin{equation*}
\mathbf{H}_{k} =   
	\left[ {\begin{array}{cccc}
    	  \frac{\hat{x} - x_1}{\left\|\mathbf{\hat{p}}-\mathbf{p}_{1}\right\|} & \frac{\hat{y} - y_1}{\left\|\mathbf{\hat{p}}-\mathbf{p}_{1}\right\|} & 0 & 0 \\
    	  \frac{\hat{x} - x_2}{\left\|\mathbf{\hat{p}}-\mathbf{p}_{2}\right\|} & \frac{\hat{y} - y_2}{\left\|\mathbf{\hat{p}}-\mathbf{p}_{2}\right\|} & 0 & 0 \\
    	  \cdots & \cdots & \cdots & \cdots \\
    	  \frac{\hat{x} - x_{N_A}}{\left\|\mathbf{\hat{p}}-\mathbf{p}_{N_A}\right\|} & \frac{\hat{y} - y_{N_A}}{\left\|\mathbf{\hat{p}}-\mathbf{p}_{N_A}\right\|} & 0 & 0 \\
  	\end{array} } \right]
\end{equation*}
with size $(N_A \times 4)$ and $\mathbf{\hat{p}} = [\hat{x}, \hat{y}]^T$ the current estimate of the target position.
\item The MSE of the ToA measures is given by the CRB with $\chi = 0$ in \eqref{equ:CRB_ToA} for all the links. 
\end{itemize}
The parameters and their numerical values for the algorithms and PCRB are reported in Table \ref{tab:parameters}; $E_{INIT}$ is the initial value of the energy for all the anchors transmissions. The initial allocation of energy is tuned for respecting the performance target, or $PCRB_{TARGET}$, with an outage equal to $1\%$ over all the trajectories without the application of any algorithm.

\begin{figure}[!tb]
   \centering
   \subfloat[$\sigma_w = 0.025$]{
        \includegraphics[width=.48\textwidth]{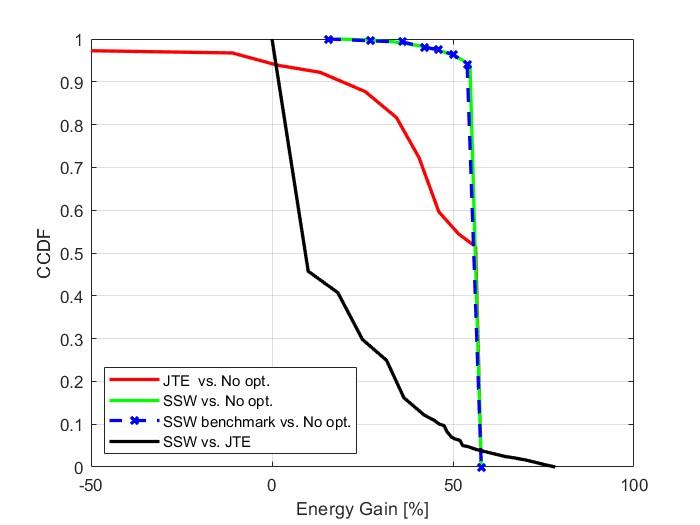}
        \label{Fig_CDF_LATG:A}
    }
    \subfloat[$\sigma_w = 1.0$]{
        \includegraphics[width=.48\textwidth]{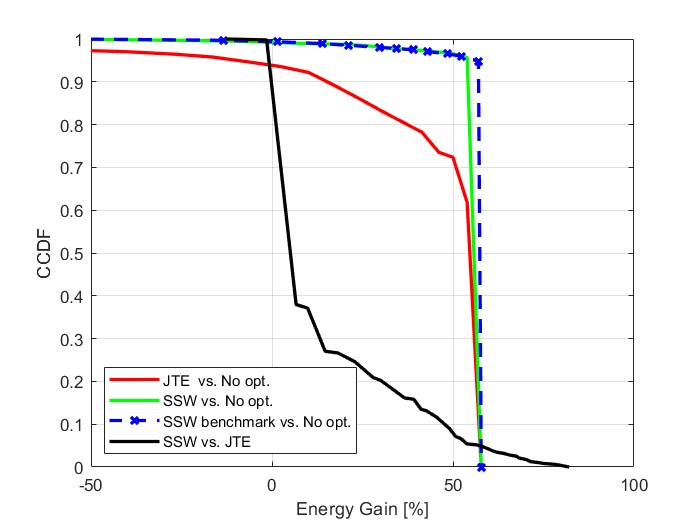}
        \label{Fig_CDF_LATG:B}
    }
 	\caption{Complementary CDF of the energy gain during the trajectories for $\sigma_w = 0.025$ (a) and $\sigma_w = 1.0$ (b) with $N_A = 4$ anchors. SSW algorithms provide a gain above $50\%$ in about $95\%$ of the cases.}
    \label{fig:Fig_CDF_LATG}
\end{figure}

In Fig. \ref{fig:Fig_CDF_PCRB}, the numerical results report the Cumulative Distribution Functions (CDF) of the PCRB w/o the application of the optimization, with the application of the two algorithms, SSW, JTE and the benchmark for two values of $\sigma_w$, representative of slow motion (root mean square of the velocity during the trajectory $v_{rms} = 0.89$ m/s) and fast motion ($v_{rms} = 47$ m/s). The corresponding Complementary CDF (CCDF) of the energy gain percentage between the solutions is reported in Fig. \ref{fig:Fig_CDF_LATG}; the energy gain is defined as the saved energy after the application of the algorithm w.r.t. energy consumed w/o the algorithms (expressed as a percentage). 
After the application of the algorithms, the initial energy allocation ($E_{INIT}$) appears redundant w.r.t. the performance target and we can observe how the optimization produces two effects: (i) the PCRB is approaching the $PCRB_{TARGET}$ since the transmitted energy by the anchors is reduced and optimized w.r.t. that objective, (ii) and an energy gain appears as a result of that reduction correspondingly. 

\begin{figure}[!tb]
   \centering
   \subfloat[$\sigma_w = 0.025$]{
        \includegraphics[width=.48\textwidth]{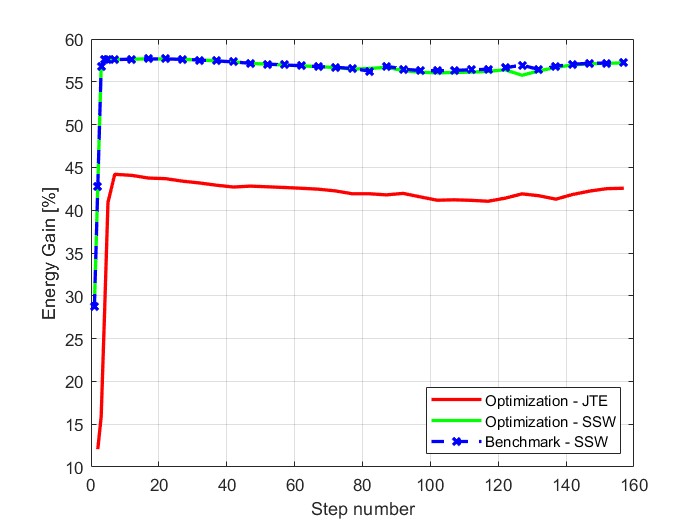}
        \label{Fig_LATG_step:A}
    }
    \subfloat[$\sigma_w = 1.0$]{
        \includegraphics[width=.48\textwidth]{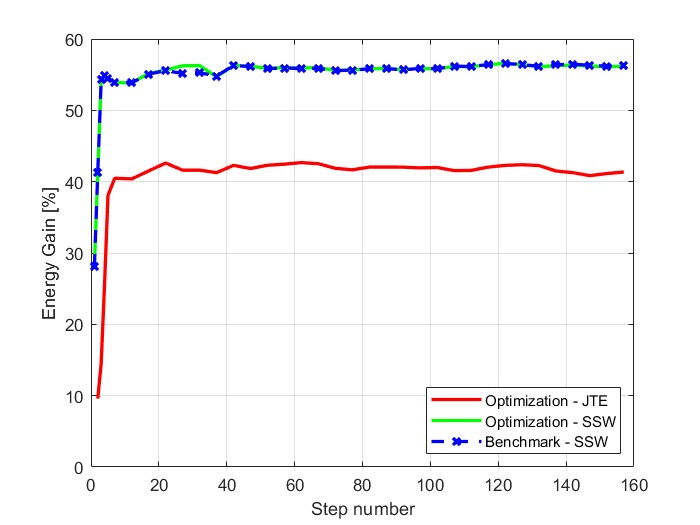}
        \label{Fig_LATG_step:B}
    }
 	\caption{Average energy gain as a function of the trajectory step for $\sigma_w = 0.025$ (a) and $\sigma_w = 1.0$ (b) as a function of the step $k$ with $N_A = 4$ anchors.}
    \label{fig:Fig_LATG_step}
\end{figure}

We can also notice that the sub-optimal JTE algorithm shows a general performance loss in the potential gain, around $20\%$, as can be observed in the curves of the energy gain as a function of the step, shown in Fig. \ref{fig:Fig_LATG_step}; it is also interesting to notice how the algorithms achieve usually the convergence fast, after a few steps, approximately $10$ in the examples. At the same time, it is also important to observe that the SSW performance overlaps perfectly the benchmark, proving that the first order approximation introduced in this paper is effective in estimating the PCRB corresponding to a change in the transmit energy.

\begin{figure}[!tb]
   \centering
   \subfloat[$\sigma_w = 0.025$]{
        \includegraphics[width=.48\textwidth]{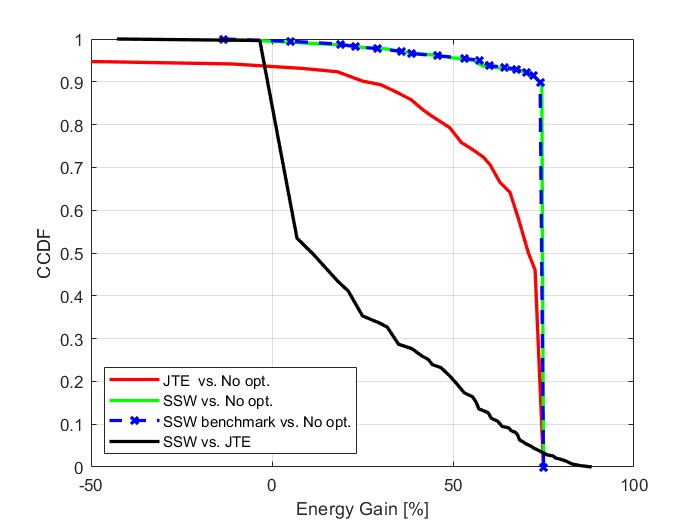}
        \label{Fig_CDF_LATG_RICE:A}
    }
    \subfloat[$\sigma_w = 1.0$]{
        \includegraphics[width=.48\textwidth]{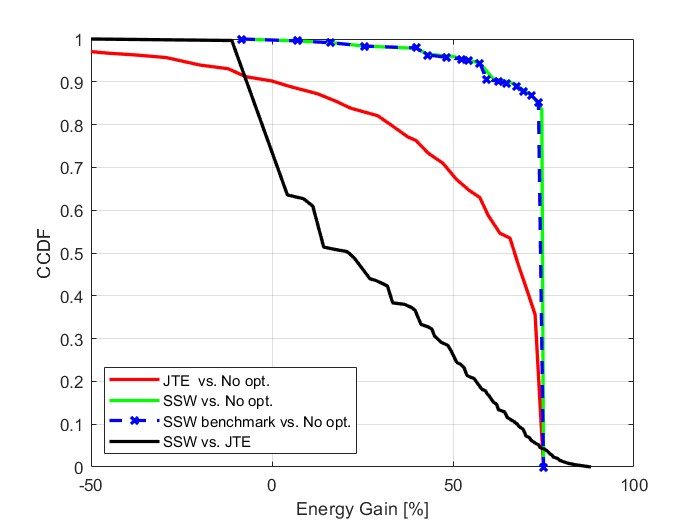}
        \label{Fig_CDF_LATG_RICE:B}
    }
 	\caption{Complementary CDF of the energy gain during the trajectories for $\sigma_w = 0.025$ (a) and $\sigma_w = 1.0$ (b) with $N_A = 4$ anchors and Rician fading with Rice factor equal to $0$ dB.}
    \label{fig:Fig_CDF_LATG_RICE}
\end{figure}

The presence of multipath fading obviously increases the absolute values of the energies necessary for respecting the accuracy constraint (including $E_{INIT}$) but not significantly the relative values and gains; Fig. \ref{fig:Fig_CDF_LATG_RICE} reports the CCDFs of the energy gains, similarly to Fig. \ref{fig:Fig_CDF_LATG}.

\begin{figure}[!tb]
   \centering
   \subfloat[Average energy gain as a function of the step]{
        \includegraphics[width=.48\textwidth]{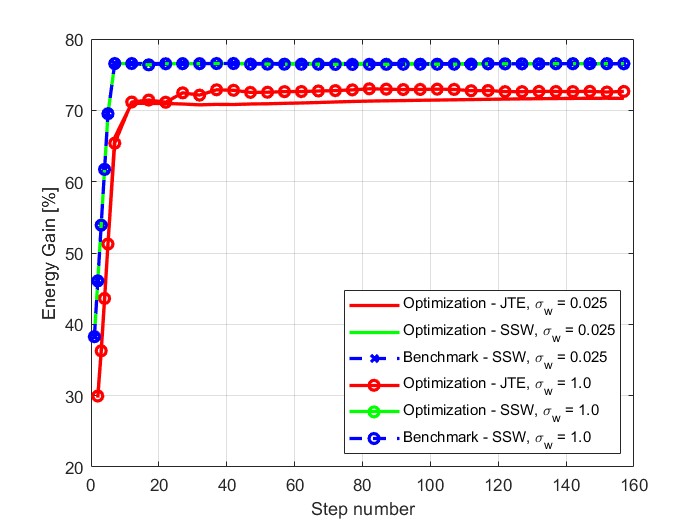}
        \label{Fig_LATG_Nb6:A}
    }
    \subfloat[Complementary CDF of the energy gain]{
        \includegraphics[width=.48\textwidth]{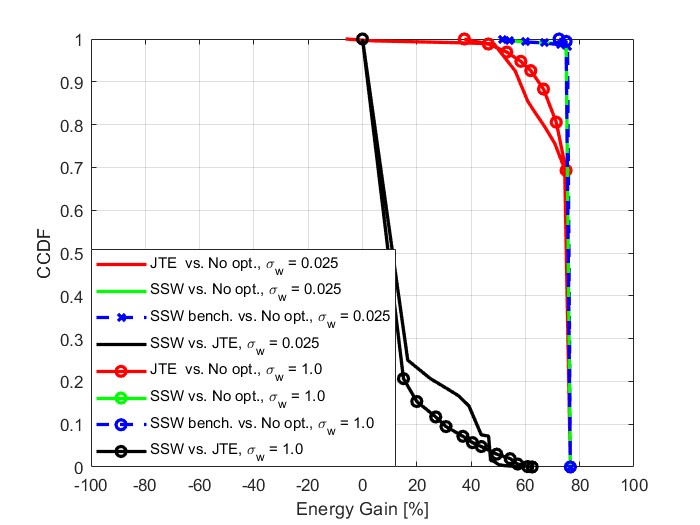}
        \label{Fig_LATG_Nb6:B}
    }
 	\caption{Average energy gain as a function of the trajectory step (a) and complementary CDF of the energy gain (b) for $\sigma_w = [0.025, 1.0]$ with $N_A = 6$ anchors.}
    \label{fig:Fig_LATG_Nb6}
\end{figure}

Increasing the number of anchors, as shown in Fig. \ref{fig:Fig_LATG_Nb6}, we observe that the algorithms gain increases (as a percentage w.r.t. absence of any strategy) under the same network and channel conditions since the number of available combinations $N_{SSW,COMB}$ for the allocation increases exponentially with $N_A$ (Sect. \ref{subsec:sliding_space}). At the same time, the difference between the SSW and JTE tends to decrease.


\section{Conclusions}\label{sec:conclusions}

In this work we have formulated and analyzed the problem of the energy minimization, constrained by an accuracy requirement in a tracking problem, in which the measures are constituted by the ToA estimates derived w.r.t. a set of known anchors. The analysis of the impact of the variation of the transmitted energy on the predicted tracking error in the context of the Posterior CRB, has revealed the role and relation of the transmission energy, the SNR and the position accuracy. After the problem formulation on the joint tracking accuracy and energy optimization, we have developed the linearization of the Posterior CRB w.r.t. energy, leading to \eqref{equ:Delta_P_qj} and \eqref{equ:discrete_pcrbvsenergy}, which are the main contribution. Then we have presented two algorithms with reduced computational complexity, able to show the applicability of the approach to realistic cases and some numerical examples of the potential advantage in terms of transmission energy savings.

\appendices
\section{Derivation of the PCRB gradient} \label{App:PCRB_gradient}
Let us consider the gradient $\nabla f_k(\mathbf{E})|_{\mathbf{E}_k}$ of $f_k(\mathbf{E}) = PCRB_{pos,k}(\mathbf{E})$ with respect to the $N_A$ energies in the vector $\mathbf{E}$, i.e.
\begin{equation}\label{equ_app:pcrbvsenergy}
\nabla f_k(\mathbf{E}) = (\frac{\partial f_k(\mathbf{E})}{\partial E_1}, \frac{\partial f_k(\mathbf{E})}{\partial E_2}, \cdots, \frac{\partial f_k(\mathbf{E})}{\partial E_{N_A}})^T.
\end{equation}
Considering that, in the 2D plane,
\begin{equation*}
f_k(\mathbf{E}) = \mathbf{P}_{k|k}(1,1) + \mathbf{P}_{k|k}(2,2),
\end{equation*}
we derive the gradient of the matrix $\mathbf{P}_{k|k}$ w.r.t. a generic element $E_j$ of $\mathbf{E}$. From \eqref{eq:Cov_matrix} and \eqref{eq:Kalman_gain}, written as
\begin{equation}
\mathbf{K}_{k} = \mathbf{P}_{k|k-1} \mathbf{H}_k^T (\mathbf{S}_{k})^{-1},
\end{equation} 
we can write
\begin{equation}\label{equ_app:pcrb_dev}
\mathbf{P}_{k|k} = \mathbf{P}_{k|k-1} - \mathbf{P}_{k|k-1} \mathbf{H}_{k}^T \mathbf{S}_{k}^{-1} \mathbf{H}_{k} \mathbf{P}_{k|k-1}^T. 
\end{equation}
The gradient of the only term $\mathbf{S}_{k}$ in \eqref{equ_app:pcrb_dev} that depends on the energy can be expressed using the well-known relation 
\begin{equation}
\frac{\partial \mathbf{S}_{k}^{-1}}{\partial E_j} = -\mathbf{S}_{k}^{-1}\frac{\partial \mathbf{S}_{k}}{\partial E_j} \mathbf{S}_{k}^{-1},
\end{equation}
and, consequently, 
\begin{align}\label{equ_app:pcrb_gradient}
\frac{\partial \mathbf{P}_{k|k}}{\partial E_j} & = - \mathbf{P}_{k|k-1} \mathbf{H}_{k}^T \frac{\partial \mathbf{S}_{k}^{-1}}{\partial E_j} \mathbf{H}_{k} \mathbf{P}_{k|k-1}^T \nonumber \\ 
		& =  \mathbf{P}_{k|k-1} \mathbf{H}_{k}^T \mathbf{S}_{k}^{-1} \frac{\partial \mathbf{S}_{k}}{\partial E_j} \mathbf{S}_{k}^{-1} \mathbf{H}_{k} \mathbf{P}_{k|k-1}^T  \nonumber \\
		& = \mathbf{K}_{k} \frac{\partial \mathbf{S}_{k}}{\partial E_j} \mathbf{K}_{k}.
\end{align} 
Now, from \eqref{equ:S}, it is easy to realize that
\begin{equation*}
\frac{\partial \mathbf{S}_{k}}{\partial E_j} = \frac{\partial \mathbf{R}_{k}}{\partial E_j},
\end{equation*}
where $\mathbf{R}_{k}$ is the diagonal measure covariance matrix (assumed independent each other), filled with the ToA-based distances CRBs defined in \eqref{equ:CRB_ToA}, considered at time $k$. Therefore, as $E_j$ is only in the j-th element of the diagonal of $\mathbf{R}_{k}$, we get  
\begin{equation}\label{equ_app:dS_dE}
\frac{\partial \mathbf{S}_{k}}{\partial E_j} =   
	\left[ {\begin{array}{ccccc}
    	  0      & 0      & \cdots                                     & 0      & 0 \\
    	  0      & 0      & \cdots                                     & 0      & 0 \\
    	  \cdots & \cdots & \frac{\partial CRB_{dis,j,k}}{\partial E_j}  & \cdots & 0 \\
    	  \cdots & \cdots & \cdots                                     & \cdots & 0 \\
    	  0      & 0      & 0                                          & \cdots & 0 \\
  	\end{array} } \right],
\end{equation}
and, replaced in \eqref{equ_app:pcrb_gradient},
\begin{equation*}
\frac{\partial \mathbf{P}_{k|k}}{\partial E_j} = \frac{\partial CRB_{dis,j,k}}{\partial E_j} \left[ {\begin{array}{ccc}
    	  K_k(1,j)^2  &  \cdots    & \cdots    \\
    	  \cdots      & K_k(2,j)^2 & \cdots    \\
    	  \cdots      & \cdots     & \cdots    \\
    	  \cdots      & \cdots     & \cdots    \\
  	\end{array} } \right].
\end{equation*}
Finally, from \eqref{equ_app:pcrb_gradient}, we obtain the final result
\begin{equation}\label{equ_app:pcrb}
\frac{\partial f_k(\mathbf{E})}{\partial E_j} = q_{j,k} = (K_k(1,j)^2 + K_k(2,j)^2) \frac{\partial CRB_{dis,j,k}}{\partial E_{j}}.
\end{equation}
When the $CRB_{dis,j,k}$ is inversely proportional to the energy $E_{j}$, as in our ToA-based model \eqref{equ:CRB_ToA}, we get, for the derivative evaluated at time $k$,
\begin{equation}\label{equ_app:q}
q_{j,k} = - \frac{(K_k(1,j)^2 + K_k(2,j)^2) \cdot CRB_{dis,j,k}}{E_{j,k}}.
\end{equation}
%

%
%




\ifCLASSOPTIONcaptionsoff
  \newpage
\fi

\bibliographystyle{IEEEtran}
%
\bibliography{IEEEabrv, Bibliography}

@article{3GPPtr38857,
title={{3rd Generation Partnership Project;
Technical Specification Group Services and System Aspects;
Study on NR Positioning Enhancements, 
(Release 17)}},
journal={3GPP Tech. Report},
author={{TR 38.857}},
volume={17},
number={},
year={2021},
publisher={http://www.3gpp.org/}
}

@article{3GPPtr22804,
title={{3rd Generation Partnership Project;
Technical Specification Group Services and System Aspects;
Study on Communication for Automation in Vertical Domains;
(Release 16)}},
author={{TR 22.804}},
journal={3GPP Tech. Report},
volume={16},
number={},
year={2020},
publisher={http://www.3gpp.org/}
}

@article{brown2018ultra,
  title={{Ultra-reliable low-latency 5G for industrial automation}},
  author={Brown, Gabriel},
  journal={Technol. Rep. Qualcomm},
  volume={2},
  pages={5206-5394},
  year={2018}
}

@article{Moe.Z.Win2016,
author = {Han, Yanjun and Shen, Yuan and Zhang, Xiao-Ping and Win, Moe Z. and Meng, Huadong},
doi = {10.1109/TIT.2015.2511778},
journal = {IEEE Transactions on Information Theory},
volume = {62},
pages = {1074-1075},
title = {{Performance Limits and Geometric Properties of
Array Localization}},
year = {2016}
}

@article{AnnaGuerra2018,
author = {Guerra, Anna and Guidi, Francesco and Dardari, Davide},
doi = {10.1109/TWC.2018.2840136},
issn = {1558-2248},
journal = {IEEE Transactions on Wireless Communications},
pages = {5241-5255},
title = {{Single-Anchor Localization and Orientation Performance Limits Using Massive Arrays: MIMO vs. Beamforming}},
volume = {17},
year = {2018}
}

@article{Moe.Z.Win2010,
author = {Shen, Yuan and Win, Moe Z.},
doi = {10.1109/TIT.2010.2060110},
issn = {1557-9654},
journal = {IEEE Transactions on Information Theory},
pages = {4956-4980},
publisher = {IEEE},
title = {{Fundamental Limits of Wideband Localization— Part I: A General Framework}},
volume = {56},
year = {2010}
}

@article{RobertM.TaylorJr2003,
author = {Taylor Jr, Robert M. and Flanagan, Brian P. and Uber, John A.},
doi = {10.1109/ICASSP.2003.1201771},
issn = {1520-6149},
journal = {2003 IEEE International Conference on Acoustics, Speech, and Signal Processing, 2003. Proceedings. (ICASSP '03)},
title = {{Computing the recursive posterior Cramer-Rao bound for a nonlinear nonstationary system}},
year = {2003}
}

@INPROCEEDINGS{7346955,
  author={Prince, Gregary B. and Little, Thomas D.C.},
  booktitle={2015 International Conference on Indoor Positioning and Indoor Navigation (IPIN)}, 
  title={{Latency Constrained Device Positioning Using a Visible Light Communication Two-Phase Received Signal Strength - Angle of Arrival Algorithm}}, 
  year={2015},
  volume={},
  number={},
  pages={1-7},
  doi={10.1109/IPIN.2015.7346955}
}

@inproceedings{Haojun_2019,
author = {Ai, Haojun and Zhang, Sheng and Tang, Kaifeng and Li, Ning and Huang, Weiyi and Wang, Yifeng},
year = {2019},
pages = {58-72},
title = {{Robust Low-Latency Indoor Localization Using Bluetooth Low Energy}},
booktitle={Proceedings of the ION 2019 Pacific PNT Meeting}, 
doi = {10.33012/2019.16793}
}

@book{Shalom_2002,
  title     = {{Estimation with Applications to Tracking and Navigation: Theory, Algorithms and Software}},
  author    = {Bar-Shalom, Yaakov and Li, X.-Rong and Kirubarajan, Thiagalingam},
  year      = {2002},
  publisher = {John Wiley and Sons},
  isbn = {9780471416555}
}

@ARTICLE{10127621,
  author={Morselli, Flavio and Modarres Razavi, Sara and Win, Moe Z. and Conti, Andrea},
  journal={IEEE Transactions on Wireless Communications}, 
  title={{Soft Information-Based Localization for 5G Networks and Beyond}}, 
  year={2023},
  volume={22},
  number={12},
  pages={9923-9938}
}

@ARTICLE{9522056,
  author={O’Lone, Christopher E. and Dhillon, Harpreet S. and Buehrer, R. Michael},
  journal={IEEE Transactions on Wireless Communications}, 
  title={{Characterizing the First-Arriving Multipath Component in 5G Millimeter Wave Networks: TOA, AOA, and Non-Line-of-Sight Bias}}, 
  year={2022},
  volume={21},
  number={3},
  pages={1602-1620}
}

@ARTICLE{6196144,
  author={Demigha, Oualid and Hidouci, Walid-Khaled and Ahmed, Toufik},
  journal={IEEE Communications Surveys and Tutorials}, 
  title={On Energy Efficiency in Collaborative Target Tracking in Wireless Sensor Network: A Review}, 
  year={2013},
  volume={15},
  number={3},
  pages={1210-1222},
  doi={10.1109/SURV.2012.042512.00030}
}

@article{PANG2021932,
title = {A new energy efficient management approach for wireless sensor networks in target tracking},
journal = {Defence Technology},
volume = {17},
number = {3},
pages = {932-947},
year = {2021},
issn = {2214-9147},
doi = {https://doi.org/10.1016/j.dt.2020.05.022},
author = {Ce Pang and Gong-guo Xu and Gan-lin Shan and Yun-pu Zhang},
}

@INPROCEEDINGS{8269035,
  author={Wang, Donglin and Fattouche, Michel and Ghannouchi, Fadhel M.},
  booktitle={2017 IEEE Globecom Workshops (GC Wkshps)}, 
  title={Bounds of mmWave-Based Ranging and Positioning in Multipath Channels}, 
  year={2017},
  volume={},
  number={},
  pages={1-6},
  doi={10.1109/GLOCOMW.2017.8269035}
}

@ARTICLE{9119156,
  author={Zhang, Haowei and Zong, Binfeng and Xie, Junwei},
  journal={IEEE Transactions on Vehicular Technology}, 
  title={Power and Bandwidth Allocation for Multi-Target Tracking in Collocated {MIMO} Radar}, 
  year={2020},
  volume={69},
  number={9},
  pages={9795-9806},
  doi={10.1109/TVT.2020.3002899}}

@article{3GPPtr21917,
title={{3rd Generation Partnership Project;
Technical Specification Group Services and System Aspects;
Release 17 Description;Summary of Rel-17 Work Items;
(Release 17)}},
author={{TR 21.917}},
journal={3GPP Tech. Report},
volume={17.0.1},
number={},
year={2023},
publisher={http://www.3gpp.org/}
}

@book{Van_Trees_1968,
  title     = {{Detection, Estimation, and Modulation Theory}},
  author    = {Van Trees, Harry L.},
  year      = {1968},
  publisher = {John Wiley and Sons},
  edition = {1st}
}

@ARTICLE{6781631,
  author={Garcia, Gabriel E. and Muppirisetty, L. Srikar and Schiller, Elad M. and Wymeersch, Henk},
  journal={IEEE Transactions on Wireless Communications}, 
  title={On the Trade-Off Between Accuracy and Delay in Cooperative UWB Localization: Performance Bounds and Scaling Laws}, 
  year={2014},
  volume={13},
  number={8},
  pages={4574-4585}
}

@ARTICLE{10620268,
  author={Chen, Geng and Cheng, Lili and Zeng, Qingtian and Shen, Fei and Zhang, Yu-Dong},
  journal={IEEE Transactions on Green Communications and Networking}, 
  title={Willingness Allocation-Assisted Cooperative Localization Algorithm Based on Competitive Game for Resource-Constrained Environment}, 
  year={2025},
  volume={9},
  number={2},
  pages={498-512},
  doi={10.1109/TGCN.2024.3436535}
}

@ARTICLE{6807568,
  author={Alirezaei, Gholamreza and Reyer, Michael and Mathar, Rudolf},
  journal={IEEE Transactions on Wireless Communications}, 
  title={Optimum Power Allocation in Sensor Networks for Passive Radar Applications}, 
  year={2014},
  volume={13},
  number={6},
  pages={3222-3231},
  doi={10.1109/TWC.2014.042114.131870}
}

@ARTICLE{Hchen_10049506,
  author={Chen, Huimin and Xing, Fangyuan and Yang, Qianqian and Shu, Yuanchao and Shi, Zhiguo and Chen, Jiming and Tao, Zhen},
  journal={IEEE Transactions on Green Communications and Networking}, 
  title={A Lightweight Mobile-Anchor-Based Multi-Target Outdoor Localization Scheme Using LoRa Communication}, 
  year={2023},
  volume={7},
  number={4},
  pages={1607-1619},
  doi={10.1109/TGCN.2023.3236630}
}

@ARTICLE{Huang_10824973,
  author={Huang, Jieyu and Xie, Junwei and Yang, Ziqing and Zhang, Haowei and Li, Zhengjie and Feng, Weike},
  journal={IEEE Transactions on Vehicular Technology}, 
  title={Joint Resource Allocation Strategy for Multiple Target Tracking in Networked Collocated MIMO Radar System}, 
  year={2025},
  volume={74},
  number={5},
  pages={7201-7211},
  doi={10.1109/TVT.2025.3525642}
}

@ARTICLE{Li_9881913,
  author={Li, Zhengjie and Xie, Junwei and Liu, Weijian and Zhang, Haowei and Xiang, Houhong},
  journal={IEEE Transactions on Vehicular Technology}, 
  title={Joint Strategy of Power and Bandwidth Allocation for Multiple Maneuvering Target Tracking in Cognitive MIMO Radar With Collocated Antennas}, 
  year={2023},
  volume={72},
  number={1},
  pages={190-204},
  doi={10.1109/TVT.2022.3204939}
}

%


%








\end{document}